# A superfluid $^4$He interferometer operating near 2 K

Emile Hoskinson, Yuki Sato, and Richard Packard

*Department of Physics, University of California, Berkeley, CA 94720 USA*

**Matter-wave interferometers reveal some of the most fascinating phenomena of the quantum world[1]. Phase shifts due to rotation (the Sagnac effect) for neutrons[2], free atoms[3] and[4,5] superfluid $^3$He reveal the connection of matter waves to a non-rotating inertial frame. In addition, phase shifts in electron waves due to magnetic vector potentials (the Aharonov-Bohm effect[6]) show that physical states can be modified in the absence of classical forces. We report here the observation of interference induced by the Earth's rotation in superfluid $^4$He at 2 K, a temperature 2000 times higher than previously achieved with $^3$He. This interferometer, an analog of a dc-SQUID, employs a recently reported phenomenon wherein superfluid $^4$He exhibits quantum oscillations in an array of sub-micron apertures[7,8]. We find that the interference pattern persists not only when the aperture array current-phase relation is a sinusoidal function characteristic of the Josephson effect, but also at lower temperatures where it is linear and oscillations occur by phase slips[9,10]. The modest requirements for the interferometer (2 K cryogenics and fabrication of apertures at the level of 100nm) and its potential resolution suggest that, when engineering challenges such as vibration isolation are met, superfluid $^4$He interferometers could become important scientific probes.**

Superfluid $^4$He is described by a macroscopic wave function which includes a phase $\phi$ that can vary in time and space. Our new interferometer is based on a recently reported phenomenon wherein an array of 4225 ($65^2$) small apertures coupling together two reservoirs of superfluid $^4$He exhibits fluid oscillations at the Josephson frequency.

The superfluid current $I$ through an aperture array is a function of the fluid phase difference $\Delta\phi$ between one side of the array and the other. This phase difference, and the corresponding fluid flow, evolves in time according to the Josephson-Anderson phase evolution equation $d\Delta\phi/dt = -\Delta\mu/\hbar$. Here $\hbar$ is Plank's constant $h$ divided by $2\pi$. The chemical potential difference $\Delta\mu$ across the array includes both the temperature drop, $\Delta T$, and pressure drop, $\Delta P$, across the array: $\Delta\mu = m_4(\Delta P/\rho - s\Delta T)$. Here $m_4$ is the mass of a $^4$He atom, $\rho$ is the fluid mass density, and $s$ its entropy per unit mass. A constant chemical potential difference applied across the array gives rise to flow through it that oscillates at the Josephson frequency $f_J = \Delta\mu/h$. Depending on the proximity to the superfluid transition temperature, 2.17 K, the oscillations are due either to a dc-Josephson type current-phase relation, $I \propto \sin\phi$, or to a linear current-phase relation with $2\pi$ phase slips. In either case, as long as the temperature is within about 10mK of the transition, the oscillations are synchronous throughout the array.

Our interferometer, which is equivalent to a dc Superconducting Quantum Interference Device (dc-SQUID), is shown schematically in Fig. 1a. We place two arrays within tubes filled with $^4$He that form an interferometer "loop". Each array contains 4225 apertures, nominally 90 nm in diameter, spaced on a 3μm square lattice in a 50nm thick silicon nitride membrane. The enclosed "sense area" of the loop is nominally 10 cm$^2$. The arrays are positioned equidistant from a diaphragm displacement transducer that functions as a microphone to detect the oscillating currents. Using a combination of electrostatic forces applied to the diaphragm and power applied to a heater just below the diaphragm we create and control chemical potential differences. This $\Delta\mu$ is felt equally by both arrays, and current oscillates through each at a constant Josephson frequency (typically near 700Hz) for periods on the order of 15 seconds. The microphone motion is a linear superposition of the oscillations in the two arrays.

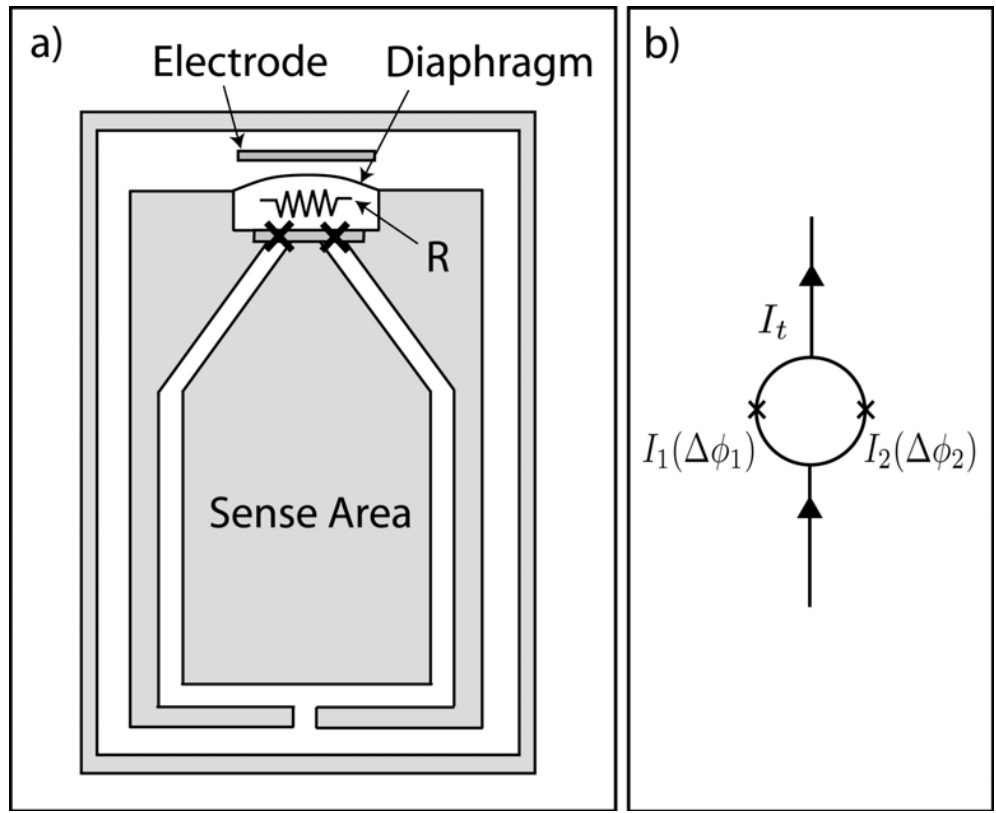

Figure1. Superfluid $^4$He interferometer schematic and equivalent circuit. a. A sketch of the $^4$He interferometer. The x's indicate the position of the two aperture arrays. The unshaded regions are filled with superfluid $^4$He. The upper chamber is closed on the top by a flexible metalized diaphragm which serves both as a microphone to detect the Josephson oscillations and also as a pressure pump to maintain chemical potential differentials across the arrays. Pressure is created by application of an electrostatic force between the diaphragm and electrode. The resistor R is a heater which can contribute to the chemical potential differential. Above the electrode is a superconducting pancake coil (not shown) which is part of a SQUID-based sensor[17] used to detect the motion of the diaphragm. The interferometer is inside a can (outer shaded border) with the superfluid $^4$He inside pressurized to roughly half an atmosphere. The can is immersed in a conventional dewar filled with helium whose temperature T is feedback regulated[18] with a stability of ~50nK just below its superfluid transition temperature $T_\lambda$ = 2.17K. Since the can is pressurized, the value of $T_\lambda$ for the liquid inside it is slightly lower than that of the bath -- this makes it easier to stabilize T very close to the $T_\lambda$ of the can. A sketch of the equivalent interferometer topology, emphasizing the analogy with the dc-SQUID, is shown in part b.

If there is a well-defined phase difference $\Delta\phi \equiv \Delta\phi_1 - \Delta\phi_2$ between the arrays, then by superposition the signal amplitude of the total current $I_t = I_1 + I_2$ detected at the microphone can be written as

$$I_m = 2I_c \left|\cos\frac{\Delta\phi}{2}\right| \qquad 1.$$

Here the two arrays are assumed to have equal flow oscillation amplitudes $I_c$. This is the typical behavior of an interferometer: a phase difference between two paths modulates the combined signal. The essence of any interferometer experiment is to introduce a $\Delta\phi$ by external means. In this work we do so via a rotation-induced Sagnac effect.

The wave function of the superfluid in the interferometer is single valued. This implies that, integrating around the closed loop of the interferometer, $\oint \vec{\nabla}\phi \cdot d\vec{\ell} = 2\pi n$, where $n$ is an integer. In this integral there is a contribution of $\Delta\phi_1$ from the left aperture array and $-\Delta\phi_2$ from the right one.

Phase gradients correspond to fluid flow and, in the tubes connecting the arrays, $\nabla\phi = m_4 v_s / \hbar$, where $v_s$ is the superfluid velocity. If the interferometer is rotating with angular velocity $\vec{\Omega}$, the fluid in the connecting tubes moves with it and gives rise to a contribution to the loop phase integral, $4\pi \vec{\Omega} \cdot \vec{A}/\kappa_4$, where $\vec{A}$ is the loop area vector and $\kappa_4 = h/m_4$ is the $^4$He quantum of circulation. The integral therefore yields

$$\Delta\phi \equiv \Delta\phi_1 - \Delta\phi_2 = 4\pi \frac{\vec{\Omega} \cdot \vec{A}}{\kappa_4} + 2\pi n + \phi_b. \qquad 2.$$

Here $\phi_b$ represents any other fixed current bias in the loop. This equation shows how an angular velocity gives rise to a phase shift between the arms of the interferometer. This shift, combined with Eq. 1, predicts that the microphone signal should be modulated as

$$I_m = 2I_c \left|\cos\left(\frac{\pi 2\vec{\Omega} \cdot \vec{A}}{\kappa_4} + \phi_b + 2\pi n\right)\right| \qquad 3.$$

When we apply chemical potential differences across the interferometer loop, the microphone output easily resolves the resulting mass current oscillating at the Josephson frequency. We choose a frequency near 700Hz because it lies in a spectral window that is away from acoustic nuisance signals picked up by the displacement sensor. The amplitude of this acoustic signal is our main experimental probe of the system.

Our interferometer rotates with the Earth. The area vector $\vec{A}$, which is normal to the plane of the interferometer, lies in the horizontal plane of our laboratory. Our experiment consists of reorienting the cryostat and the interferometer within it about a vertical axis in the laboratory, thus changing the scalar product between $\vec{A}$ and the Earth's rotation vector $\vec{\Omega}_E$. We excite Josephson oscillations by establishing chemical potentials differentials across the array by applying bias voltages to the diaphragm and electric power to a heater within the upper cell. The measurement time is restricted by the time it takes for the diaphragm to be stretched to some limiting value. This time varies inversely with the critical velocity at which the fluid flows (in the phase slip regime). Since the critical velocity increases at lower temperatures, we cannot record meaningful data in the present apparatus below about $T_\lambda - T \approx 20 mK$. For each orientation we record the amplitude of the Josephson frequency oscillations measured by the microphone. Figure 2 shows a plot of this amplitude as a function of $2\vec{\Omega}_E \bullet \vec{A}/\kappa_4$.

This is the central result of this experiment. The family of curves dramatically displays the periodic modulation predicted by Eq. 3, the pattern characteristic of a double path interferometer.

Each curve in Fig. 2 represents one fixed temperature. The absolute current calibration (a multiplicative factor) was obtained using a procedure developed for single array measurements which works equally well for a double array configuration. The curves have been shifted horizontally so that the minima align. This compensates for bias currents that are not yet completely under our control, zeroing $\phi_b$ in equations 2 and 3.

The fact that the modulation curves do not drop all the way to zero at the minima is due to unequal oscillation amplitudes of the two arrays, yielding incomplete destructive interference. A more general expression than Eq. 1 for the microphone modulation when the oscillation amplitudes $I_{c1}$ and $I_{c2}$ for the two arrays are unequal is

$$I_m = 2I_{c3}\left[\cos^2\left(\frac{\Delta\phi}{2}\right) + \gamma^2 \sin^2\left(\frac{\Delta\phi}{2}\right)\right]^{\frac{1}{2}} \qquad 4,$$

where $I_{c3} = (I_{c1} + I_{c2})/2$ and $\gamma = (I_{c1} - I_{c2})/(I_{c1} + I_{c2})$. The asymmetry parameter $\gamma$ varies from zero, where $I_{c1} = I_{c2}$ permits 100% modulation, to unity, when one array is completely blocked and no modulation exists.

For our proof-of-principle demonstration the interferometer area was chosen to be sufficiently large so that the Earth's rotation would create a Sagnac shift $\Delta\phi$ of at least $2\pi$. We also required the area to be sufficiently small that random rotation signals associated with rocking motion of the cryostat support structure would not swamp the output.

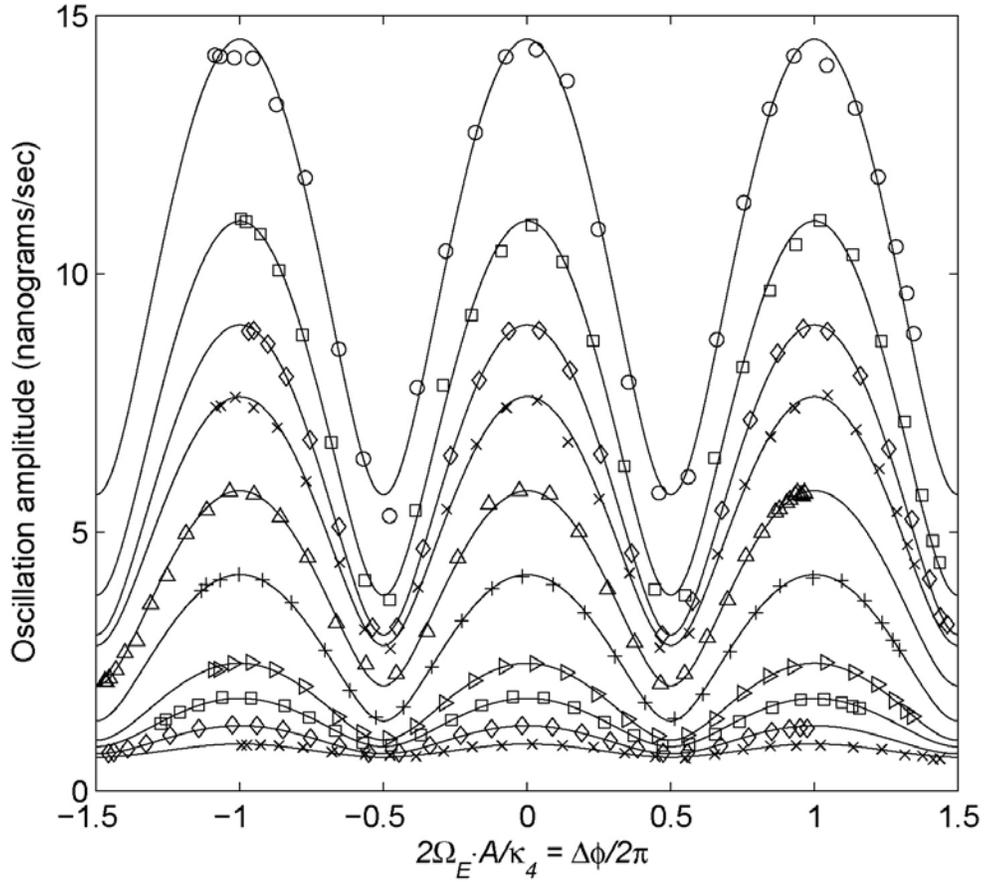

Figure 2. Modulation of the amplitude of Josephson frequency oscillations as a function of rotation flux in a superfluid $^4$He matter-wave interferometer. The Earth is the source of rotation flux. The flux magnitude is varied by reorienting the interferometer with respect to the North-South axis of the Earth. The measured data is shown by the symbols. The solid lines are fits of the data to Eq. 4, strikingly showing that the measured data follows the predicted expression throughout the temperature range investigated. From top to bottom, the modulation curves were taken at temperatures $T_\lambda - T = 12, 7.0, 4.0, 3.0, 2.0, 1.5, 0.9, 0.6, 0.4, 0.3$ mK

The period of modulation for each of the measured curves in Fig. 2 accurately corresponds to the value expected from Eq. 2 with an enclosed loop area of 10.6 cm$^2$. The validity of Eq. 4 is further strengthened by its accurate reproduction of the shape of the modulation data. From each fit of Eq. 4 to the data, we determine the maximum current $I_m$ and the asymmetry parameter $\gamma$. The steady decrease of $I_m$ as T approaches $T_\lambda$, clearly evident from the decreasing maximum value of the curves in Fig. 2, is expected as the superfluid density decreases and the superfluid healing length increases. We find that $\gamma$ decreases from 0.8 very near $T_\lambda$ to a temperature independent value ~ 0.4 below $T_\lambda - T$ ~ 2 mK. This corresponds to the approximate temperature where the array $I(\Delta\phi)$ functions change from sinusoidal to linear. It is important to emphasize that the interference is still distinct and described by Eq. 4 well into the linear current-phase regime.

Sagnac interference in a superfluid double aperture interferometer has been reported previously for $^3$He, a superfluid that only exists below about $10^{-3}$K. Our observation here of quantum interference in $^4$He at 2000 times higher temperature renders this phenomenon much more accessible. In addition, the fact that the interference persists when the current-phase relation is linear suggests that the phenomenon is robust. Although many technical hurdles remain, these two new features move the phenomenon of superfluid interference from a difficult to access laboratory curiosity toward a practical laboratory tool for fundamental and applied research.

The instrumental limit for phase shift measurements of this particular interferometer can be estimated by multiplying the inverse slope of fig. 2 by the smallest detectable current, $\delta I_{min} = \rho a \omega \delta x_{min}$ where $a = 0.5\, cm^2$ is the diaphragm detector area, $\omega$ is the Josephson frequency of the measurement and $\delta x_{min} \approx 3 \times 10^{-15} m$ is the smallest displacement that can be detected in a one Hertz bandwidth. Thus,

$$\delta\phi_{min} = \frac{d\phi}{dI} \rho a \omega \delta x_{min} \qquad\qquad 5.$$

For $\omega = 2\pi \times 700\, Hz$, the 12mK curve in Fig. 2 yields $\delta\phi_{min} \approx 3 \times 10^{-2}$ rad in a one Hertz bandwidth. With the 10 cm$^2$ loop area of this proof-of-principle device, this translates into a minimum observable angular velocity change $\delta\Omega_{min} \approx 2 \times 10^{-7}$ rad/sec in a one Hertz bandwidth. Although analysis' of the thermal noise limits of related devices have been performed[11,12], the corresponding limit for our device is not yet known. We are currently limited by the random rocking motion of our air spring cryostat support system, which creates nuisance rotation signals approximately one order of magnitude greater than the electronic background. This corresponds to a tilt displacement at one end of the 1.5 m long cryostat of less than 1 μm. This is the origin of the small scatter in Fig. 2. Utilization of the superfluid interferometer as a geodetic gyroscope requires a more rigid structure placed in a quiet environment[13]. Alternatively, as a device for inertial navigation in a noisy environment, one will need to employ feedback to lock the device at a particular bias point while having a slew rate high enough to track nuisance signals. In contrast, as a detector for non-rotation induced phase shifts, the loop can be configured with two turns of opposite helicity to enclose zero rotation flux while performing a differential measurement on the phenomenon of interest.

Configured as a gyroscope the angular velocity sensitivity of the interferometer improves with several variables including the number of apertures, the number of "turns" in the interferometer loop, and the enclosed area. Based on Eq. 3 and our measured microphone noise, if the aperture array contains a factor of ten more apertures, if the loop diameter is increased by a factor of 4 and if the loop geometry is reconfigured as a 10 turn helix, the extrapolated resolution of this interferometer surpasses the reported resolution of the best optical[14] and cold-atom Sagnac gyroscopes. Such a design, if anchored to a sufficiently rigid platform, would provide a useful complement to current Earth monitoring systems based on VLBI[15].

An interferometer with multiple turns can also be used to settle a debate on the predicted existence of an Aharonov-Bohm (AB) interference in neutral matter[16]. An intriguing aspect of the latter experiment is that previous analyses of AB effects assume that the particles that interfere actually traverse the spatial region where the electromagnetic potentials exist. In the superfluid interferometer the atoms themselves do not move throughout the entire geometry within the time scale of a Josephson period. A positive detection of the AB effect in neutral superfluid matter would imply that the interference comes from the existence of the macroscopic entangled state rather than from the evolution of individual propagating matter wave packets.

The interferometer discussed here functioned as described on its first cooldown, using a pair of aperture arrays that had been stored for over five years on a laboratory shelf. This suggests that the relevant phenomenon is robust and that the technology can be substantially improved in future devices. Since the operating temperature of the interferometer is near 2 K, a regime that can be attained with a mechanical cryocoolers, this type of interferometer can be accessible to investigators without substantial cryogenic experience. We believe that experiments using the superfluid $^4$He interferometer will provide a new window to observe nature and further understand the subtleties of the quantum world.

## Acknowledgements

We acknowledge useful conversations with Dung-Hai Lee and Raymond Chiao. We thank Inseob Hahn for the use of the high resolution thermometer. EH acknowledges the support of the CNRS – CRTBT during the preparation of the manuscript. This work was supported in part by the NSF and by NASA.